\documentclass{emulateapj}
\usepackage{natbib}
\usepackage{color}

\shorttitle{Radio Continuum Sizes of the Nuclear Disks of Arp 220}
\shortauthors{Barcos-Mu\~noz et al.}

\begin{document}
\title{High-Resolution Radio Continuum Measurements of the Nuclear Disks of Arp 220}

\author{L. Barcos-Mu\~noz\altaffilmark{1}}
\email{ldb7et@virginia.edu}
\author{A. K. Leroy \altaffilmark{2}}
\author{A. S. Evans \altaffilmark{1,2}}
\author{G. C. Privon \altaffilmark{1}}
\author{L. Armus \altaffilmark{3}}
\author{J. Condon \altaffilmark{2}}
\author{J. M. Mazzarella \altaffilmark{4}}
\author{D. S. Meier \altaffilmark{5,6}}
\author{E. Momjian \altaffilmark{6}}
\author{E. J. Murphy \altaffilmark{4}}
\author{J. Ott \altaffilmark{6}}
\author{A. Reichardt \altaffilmark{2,7}}
\author{K. Sakamoto \altaffilmark{8}}
\author{D. B. Sanders \altaffilmark{9}}
\author{E. Schinnerer \altaffilmark{10}}
\author{S. Stierwalt \altaffilmark{1}}
\author{J. A. Surace \altaffilmark{11}}
\author{T. A. Thompson \altaffilmark{12}}
\author{F. Walter \altaffilmark{10}}

\altaffiltext{1}{Department of Astronomy, University of Virginia, 530 McCormick Road, Charlottesville, VA 22904, USA}
\altaffiltext{2}{National Radio Astronomy Observatory, 520 Edgemont Road, Charlottesville, VA 22904, USA}
\altaffiltext{3}{Spitzer Science Center, California Institute of Technology, MC 220-6, 1200 East California Boulevard, Pasadena, CA 91125, USA}
\altaffiltext{4}{Infrared Processing and Analysis Center, MS 100-22, California Institute of Technology, Pasadena, CA 91125, USA}
\altaffiltext{5}{New Mexico Institute of Mining and Technology, 801 Leroy Place, Socorro, NM 87801, USA}
\altaffiltext{6}{National Radio Astronomy Observatory, P.O. Box O, 1003 Lopezville Road, Socorro, NM 87801, USA}
\altaffiltext{7}{NRAO NSF-sponsored REU student}
\altaffiltext{8}{Academia Sinica, Institute of Astronomy and Astrophysics, P.O. Box 23-141, Taipei 10617, Taiwan}
\altaffiltext{9}{Institute for Astronomy, University of Hawaii, 2680 Woodlawn Dr., Honolulu, HI 96816, USA}
\altaffiltext{10}{Max-Planck-Institut f\"{u}r Astronomie, K\"{o}nigstuhl 17, D-69117 Heidelberg, Germany}
\altaffiltext{11}{Spitzer Science Center, MS 314-6, California Institute of Technology, Pasadena, CA 91125, USA}
\altaffiltext{12}{Department of Astronomy and Center for Cosmology \& Astro-Particle Physics, The Ohio State University, Columbus, OH 43210, USA}

\begin{abstract}

We present new Karl G. Jansky Very Large Array radio continuum images
of the nuclei of Arp 220, the nearest ultra-luminous infrared
galaxy. These new images have both the angular resolution to study the
detailed morphologies of the two nuclei that power the galaxy merger
and sensitivity to a wide range of spatial scales. At 33 GHz, we
achieve a resolution of 0\farcs081~$\times$~0\farcs063 ($\mathrm{29.9
 \times 23.3~pc}$) and resolve the radio emission surrounding both
nuclei. We conclude from the decomposition of the radio spectral
energy distribution that a majority of the 33~GHz emission is
synchrotron radiation. The spatial distributions of radio emission in
both nuclei are well-described by exponential profiles. These have
deconvolved half-light radii (${\rm R_{50d}}$) of $51$ and $35$~pc for the
eastern and western nuclei, respectively, and they match the number
density profile of radio supernovae observed with very long baseline
interferometry. This similarity might be due to the fast cooling of
cosmic rays electrons caused by the presence of a strong ($\sim$ mG)
magnetic field in this system. We estimate extremely high
molecular gas surface densities of $2.2^{+2.1}_{-1.0} \times 10^5$
(east) and $4.5^{+4.5}_{-1.9} \times 10^5$ (west)~M$_\odot$~pc$^{-2}$,
corresponding to total hydrogen column densities of N$_{\rm H}$ = $2.7^{+2.7}_{-1.2}
\times 10^{25}$ (east) and $5.6^{+5.5}_{-2.4} \times
10^{25}$~cm$^{-2}$ (west). The implied gas volume densities are
similarly high, ${\rm n_{H_{_2}} \sim 3.8^{+3.8}_{-1.6} 10^4}$ (east)
and $\sim 11^{+12}_{-4.5} \times 10^4$~cm$^{-3}$ (west). We also
estimate very high luminosity surface densities of
$\mathrm{\Sigma_{IR} \sim 4.2^{+1.6}_{-0.7} \times 10^{13}}$~(east) and
$\mathrm{\Sigma_{IR} \sim 9.7^{+3.7}_{-2.4} \times
10^{13}~(west)~L_{\odot}~kpc^{-2}}$, and star formation rate surface
densities of $\mathrm{\Sigma_{SFR} \sim 10^{3.7\pm0.1}}$~(east) and
$\mathrm{\Sigma_{SFR} \sim
10^{4.1\pm0.1}~(west)~M_{\odot}~yr^{-1}~kpc^{-2}}$. These values,
especially for the western nucleus are, to our
knowledge, the highest luminosity surface densities and star formation
rate surface densities measured for any star-forming system. Despite these
high values, the nuclei appear to lie below the dusty Eddington limit in which radiation 
pressure is balanced only by self-gravity. The small measured sizes also imply that at
wavelengths shorter than $\lambda =1$~mm, dust absorption effects must
play an important role in the observed light distribution while below
5 GHz free-free absorption contributes substantial opacity.
According to these calculations, the nuclei of Arp 220 are only
transparent in the frequency range $\sim$ 5 to 350 GHz. Our
results offer no clear evidence that an active galactic nucleus
dominates the emission from either nucleus at 33 GHz.

\end{abstract} 


\section{Introduction}
\label{sec:intro}

Starbursts induced by major mergers are among the most extreme
environments in the universe. Despite their prodigious luminosities,
local merger-driven starbursts are very compact, with most of their
large gas reservoirs concentrated in dusty regions a few hundred
parsecs, or less, in size \citep[e.g.,][]{D&S98}. Measuring the
compactness of these starbursts is critical to understanding these
galaxies \citep[e.g.,][]{Soi99,Soi00,Sak08,DiSa13}. Robust size
measurements allow us to translate luminosities into key physical
quantities such as gas column density, optical depth, volumetric gas
density, and star formation rate and luminosity surface
densities. Although their luminosity renders them visible out to great
distances, the present-day rarity of major mergers means that even the
nearest ultra-luminous infrared galaxies (ULIRGs: defined as having
$\mathrm{L_{IR}[8-1000 \mu m] \geq 10^{12}~L_{\odot}}$) are relatively
distant ($>$ 70~Mpc). Thus, measuring the true extent of their active
regions requires high angular resolution. The extraordinary
extinctions present in these systems at both long (from free-free
absorption) and short (from dust opacity) wavelengths complicate the
interpretation of measurements at both wavelengths, compounding the
difficulty of measuring sizes for such systems.

Given the above considerations, radio observations at centimeter
wavelengths may be the best tool to study the deeply embedded, compact
structures at the heart of such systems
\citep[e.g.,][]{Nor88,Con91}. Radio interferometers can achieve
very high angular resolution and radio waves with $\nu \gtrsim$
5~GHz can penetrate large columns of dust and are largely unaffected
by free-free absorption. The recent upgrades to the Karl G. Jansky
Very Large Array (VLA) make it particularly well-suited for such
studies. In this paper, we make use of these new VLA capabilities to
achieve the best measurement to date of the structure of the nuclear
region of the nearest ULIRG, Arp 220.

At a luminosity distance of 77.2 Mpc, and an infrared luminosity of $\mathrm{L_{IR}[8 - 1000 \mu m] = 1.44 \times 10^{12}~L_{\odot}}$
\footnote{${\rm D_{L}}$ from NED; ${\rm L_{IR}}$ using Table 1 in \citet{S&M96} and IRAS flux densities from \citet{San03}. 
However, note that the assumption of isotropic emission that leads to this luminosity has some caveats (see Section \ref{sec:ir} and Appendix \ref{app:flux}).}, Arp 220
is the nearest ULIRG. CO and near-IR observations indicate that Arp 220 
is a gas rich merger with dynamical masses of $\mathrm{\sim 10^{9} M_{\odot}}$ within $\sim$~100 pc of 
each nucleus \citep{D&S98,Sak99, Gen01,Sak08, En11}. Arp 220 is obscured at optical through 
mid-IR wavelengths \citep{Sco98, Soi99, Haas01,Spo07,Arm07}, obstructing the direct view of the 
nuclear energy sources at these wavelengths. Observations in the frequency range where Arp 220 is optically thin have been 
able to resolve the system into two compact nuclear disks \citep{Nor88,Con91,D&S98,Sak08} and find disk 
sizes of $\sim$ 0\farcs2. However, in each case the measured sizes remain comparable to the size of the beam. 
VLBI observations at cm wavelengths by \citet{Smith98}, \citet{Lons06}, and \citet{Parra07} provide a higher 
resolution view, recovering a compact distribution of point-like sources that are proposed to be a combination 
of radio supernovae (RSNe) and supernova remnants (SNRs). However, these observations resolve out most 
of the emission from the disks. From the above, it is already clear that the disks are very compact, implying 
extraordinary volume densities and surface densities. The next step is to observe the disks in the optically 
thin frequency range with resolution high enough to clearly resolve them, and sensitivity to recover the full extent of 
the emission of the system at that frequency range. 

In this paper, we measure the structure of the Arp 220 nuclei with sensitive, high angular resolution
images obtained at 6 GHz and 33 GHz observed with the VLA. Based on the integrated spectral energy distribution 
(SED) model of Arp 220 \citep[see Figure 10b in][]{Ana00}, the total continuum flux density at 33 GHz is a 
mixture of thermal and non-thermal emission with a $\sim$ 1:2 ratio, while at 6 GHz this ratio is about 1:5. Observing 
at these two frequencies then helps us diagnose the dominant emission mechanism at radio wavelengths in Arp 220. We first report 
our observations, describe the calculations used to assess the disk structure, and then discuss the implications of 
our measurements. Throughout this paper, we adopt $\mathrm{H_{0} = 73~km~s^{-1}~Mpc^{-1}}$, 
$\mathrm{\Omega_{vacuum}=0.73}$, $\mathrm{\Omega_{matter}=0.27}$ and $\mathrm{v_{optical}~=~5,555~km~s^{-1}}$
(after correction to the CMB frame), such that 1\arcsec\ on the sky plane subtends 369 pc at the distance of Arp 220.

\section{Observations}
\label{sec:obs}

We observed Arp 220 using the VLA C (4-8 GHz) and Ka band (26.5-40 GHz) receivers, recording 
emission in 1 GHz wide windows centered at $\sim$ 4.7, 7.2, 29 and 36 GHz. We used all four VLA 
configurations with a total integration time ratio of 1:1:2:4 between D (lowest resolution), C, B, and A.
The total on-source integration time was 40 min at C band and 56 min at Ka band. We used 3C 286 
as the flux density and bandpass calibrator, and J1513+2338 and J1539+2744 as the complex gain 
calibrators at C and Ka bands, respectively. The data were obtained in multiple observing sessions
during the period 2010 August 18 to 2011 July 2, with the C and Ka band observations carried out in 
separate sessions. These observations are part of a larger project; the lower resolution (C and D 
configuration) results are presented in \citet{Leroy11} and \citet{Mur13}, and the final results for 
the complete sample will be reported in Barcos-Mu\~noz et al.~(in prep.).

We reduced the data using the Common Astronomy Software Application \citep[CASA,][]{McM07} 
package following the standard procedure for VLA data. Radio frequency interference (RFI) contaminating 
the C band was eliminated using the task \texttt{flagdata} in mode \texttt{rflag}. The RFI at Ka band was negligible. 
After calibration, we combined the data from all configurations, weighting them in proportion to their integration
time per visibility (i.e., 10:5:2:1 for D:C:B:A). We then imaged this combined data using the task \texttt{CLEAN} in 
mode \texttt{mfs} \citep{SaW94}, with \textit{Briggs} weighting setting \texttt{robust=0.5}. We combined all the data 
within each receiver band and cleaned using components with a variable spectral index ({\texttt{nterms=2}}) to 
obtain an interpolated image at an intermediate frequency (5.95 GHz for C band and 32.5 GHz for Ka band). Even 
after the initial calibration, we still observed phase and amplitude variations with time. To improve the images 
further, we iteratively self-calibrated in both phase and amplitude and applied extra flagging during 
this procedure as needed. The solutions for the amplitude self-calibration were carefully inspected and accepted 
as long as the time variations in the amplitude gains for each antenna were less than $\sim$~20\%. We worked 
mostly with these ``combined'' images at 5.95 and 32.5 GHz, but we also separately imaged the two~1~GHz 
windows at C band in order to derive a robust internal C band spectral index. 

To check our results, we imaged the 33~GHz data separately for each VLA configuration. In this test, we primarily 
applied iterative phase self-calibration. Amplitude self-calibration was applied (after $\sim 3$--$5$ iterations of 
phase self-calibration), but here we only derived normalized solutions that cannot change the observed flux. 
During this check, we also experimented with weighting the visibilities by the measured rms noise in each data set 
(using the CASA task \texttt{statwt}). These tests revealed that the combined A+B (two most extended) configurations 
recovered essentially all of the flux in the data, agreeing with the C and D data in both flux and morphology when 
convolved to matched resolution. The A configuration data alone recovered less flux than the B configuration, 
consistent with some spatial filtering at this highest resolution. Further, the overall flux recovered 
agrees with an interpolation of the integrated SED \citep[][]{Ana00}. We proceed using the full combined image with our confidence 
in the results reinforced by these tests; we verified that our fitting yields consistent results using the combined
image and the A+B configuration-only image.

The clean restoring beam for the combined images has a FWHM of 0\farcs48 $\times$ 0\farcs35 (177 $\times$ 129 pc) at 
position angle (p.a.) of -40$^{\circ}$ at 6 GHz (C band), and 0\farcs081 $\times$ 0\farcs063 (30 $\times$ 23 pc) at a 
p.a. of 65$^{\circ}$ at 32.5 GHz (Ka band). The rms noise 
measured from signal-free parts of the image is $\sim 14~\mu$Jy~beam$^{-1}$ (C) and $\sim 23~\mu$Jy~beam$^{-1}$ (Ka), which 
is within a factor of two of the expected theoretical noise. The final dynamic ranges of the images are 
$\mathrm{\sim 5.2 \times 10^{3}}$ and $\sim$ 280, for C and Ka band.

When reporting the measured flux densities from the final images, we consider 
three sources of uncertainty. First, we propagated the beam-to-beam noise (see above) and found its 
effect to be negligible at both bands. Second, we assessed the impact of the curve-of-growth technique 
used to measure the flux densities (see Section \ref{sec:calc}) by using the scatter of such a curve. This is 
also small, but larger for the individual nuclei because they are not perfectly separable. Finally, we estimated 
the uncertainty in the overall flux density calibration from the day-to-day variation of the flux density of the 
complex gain calibrator. This scatter (rms) is $\sim$ 12\% at Ka band, making the flux density calibration the 
dominant source of uncertainty at this band. At C band, the scatter in the curves of growth and the 
variation in the flux density calibration are comparable, i.e., $\sim$ 1\%. We sum all three uncertainty 
terms in quadrature and report the combined value in Table \ref{tbl-1}. Note that in addition to these uncertainties,
the absolute flux scale used at the VLA is estimated to be uncertain by $\approx 2\%$.

After reducing the data, we compared our 33 GHz image to 
VLA \citep{Nor88,Con91}, SMA \citep{Sak08} and ALMA archival images 
\citep[][and Scoville et al. {\it in prep}]{Wil14}. We found astrometric discrepancies 
of order 0\farcs1 (i.e., 1--2 beams) and traced the origin of these to the 
adopted position of our 33~GHz phase calibrator. When revised from the nominal
VLA position to the position reported in the VLBI calibrator catalog, the astrometric
agreement between our image and the other images improved to a fraction of
a beam. For reference, in our data the peak positions of the two nuclei at 33 GHz
are $\alpha_{2000} = \mathrm{15^{h}34^{m}57.291^{s} \pm 0.003^{s}} (\pm 0\farcs05)$, 
$\delta_{2000} =\mathrm{23^{\circ}30'11''.34}\pm 0\farcs04$ (east) and $\alpha_{2000} =
\mathrm{15^{h}34^{m}57.222^{s}\pm 0.002^{s}} (\pm 0\farcs03)$, $\delta_{2000} =
\mathrm{23^{\circ}30'11''.51}\pm 0\farcs03$ (west). We derive these position via a Gaussian fit
(CASA's {\tt imfit}) but they also closely coincide with the positions of the
highest intensity pixel for each nucleus. The uncertainties in the peak positions above may also be viewed
as our overall astrometric uncertainty, which we derive from the standard deviation between 
the positions of the highest intensity pixels from our 6 GHz image and our shifted 33 GHz image,
and the archival images from \citet{Nor88}, \citet{Con91}, \citet{Sak08}, \citet{Wil14} and Scoville et al. {\it in prep}.

\begin{figure*}[tbh]
\centering
\includegraphics[width=6.5in]{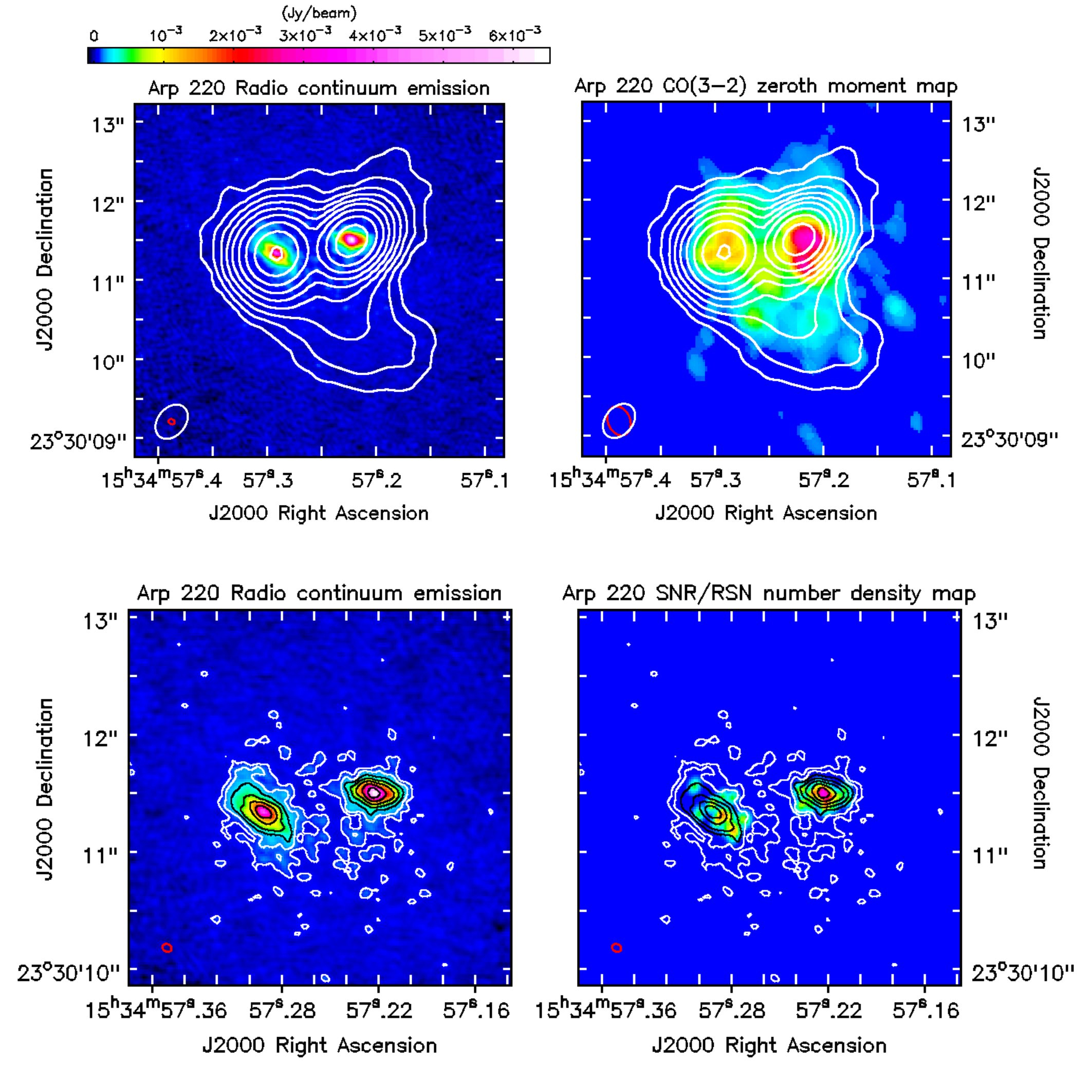}
\caption{Radio emission from Arp 220 at 6 and 33 GHz. {\em (a)} 33 GHz image (FWHM resolution of 
0\farcs081 $\times$ 0\farcs063, p.a. $\approx$ 65$^{\circ}$) with 6 GHz contours overlaid (FWHM resolution 
of 0\farcs48 $\times$ 0\farcs35, p.a. $\approx$ -40$^{\circ}$).The contours are in factor of 2 step,
with the outermost contour corresponding to $12.5\sigma$, where $\sigma$ = 14 $\mu$Jy beam$^{-1}$, and enclosing 98\% of the total flux density. In the 
lower left corner, we show the FWHM of the clean beam for the 33 GHz ({\em red}) and 6 GHz ({\em white}) image.
{\em (b)} The same 6 GHz contours overlaid on the CO(3$\to$2) integrated intensity map of \citet{Sak08} (FWHM resolution 
of 0\farcs38 $\times$ 0\farcs28, p.a. $\approx$ 23$^{\circ}$). In the lower left corner, we show the FWHM of the 
clean beam for the 33 GHz ({\em red}) and CO(3$\to$2) ({\em white}) image.{\em (c)} A 33 GHz radio continuum 
image with its contours overlaid. The contours are in factor of 2 step, with the outermost contour 
corresponding to $3\sigma$, where $\sigma$ = 23 $\mu$Jy beam$^{-1}$, and enclosing 97\% of the total flux density. {\em(d)} 33 GHz contours 
overlaid on a radio supernova (RSN) and/or supernova remnant (SNR) number density map constructed using the point sources found by \citet{Lons06}. 
We clearly resolve structure surrounding both nuclei and see a broad similarity between the radio continuum, 
gas traced by CO, and recent RSNe and/or SNRs.\label{fig1}}
\end{figure*}
 
\section{Results}
\label{sec:calc}

In Figures \ref{fig1} and \ref{fig2}, we present new VLA images of Arp 220 at 6 and 33 GHz and the 
radial profiles of each nucleus at 33 GHz. Using these new data, combined with a CO (3$\to$2) integrated intensity 
(``zeroth moment'') map from \citet{Sak08}, and positional information of point sources 
found by \citet{Lons06}, we carry out a series of calculations to determine what mechanism is
producing most of the radio emission, how radio emission traces recent star formation, and the true 
sizes and shapes of the nuclear disks. In Tables \ref{tbl-1}, \ref{tbl-2} and \ref{tbl-3} we
present the results of these calculations. 

\subsection{Integrated Flux Densities and Spectral Indices}
\label{sec:calc_flux}

\begin{deluxetable*}{cccccc}[h]
\tabletypesize{\scriptsize}
\tablecaption{Flux Densities and Spectral indices for Arp 220 
\label{tbl-1}}
\tablewidth{0pt}
\tablehead{
\colhead{Frequency} & \colhead{Total} & \multicolumn{2}{c}{East nucleus} & \multicolumn{2}{c}{West nucleus}\\ 
\colhead{(GHz)} & \colhead{(mJy)} & \colhead{Integrated (mJy)} &\colhead{Peak (mJy~beam$^{-1}$)\tablenotemark{a}}& \colhead{Integrated (mJy)} &\colhead{Peak (mJy~beam$^{-1}$)\tablenotemark{a}}}
\startdata
4.7  & 222.0 $\pm$ 1.9 & 92.4 $\pm$ 2.1 & 61.8 $\pm$ 0.5 & 114.6 $\pm$ 4.3 & 89.5 $\pm$ 0.7\\ 
7.2  & 171.4 $\pm$ 2.3 & 73.2 $\pm$ 1.4 & 36.0 $\pm$ 0.4 & 89.5 $\pm$ 1.4 & 60.4 $\pm$ 0.7 \\
5.95  & 197.6 $\pm$ 2.8 & 81.4 $\pm$ 2.8 & 49.0 $\pm$ 0.7 & 94.3 $\pm$ 1.7 & 73.3 $\pm$ 1.0\\
32.5  & 61.8 $\pm$ 7.2 & 30.1 $\pm$ 3.9 & 4.1 $\pm$ 0.5 & 33.4 $\pm$ 4.0 & 6.5 $\pm$ 0.8 \\
\cutinhead{Spectral index}
$\mathrm{\alpha_{6-33GHz}}$ & -0.69 $\pm$ 0.07 & -0.59 $\pm$ 0.08 & & -0.61 $\pm$ 0.07 \\
$\mathrm{\alpha_{4.7-7.2GHz}}$ & -0.61 $\pm$ 0.04 & -0.55 $\pm$ 0.07 &  & -0.58 $\pm$ 0.09 

\enddata
\tablecomments{For details on the calculations, see Section~\ref{sec:calc_flux}}
\tablenotetext{a}{The clean restoring beam FWHM is 0\farcs60 $\times$ 0\farcs43 at 4.2 GHz, 0\farcs38 $\times$ 0\farcs28 at 
5.95 and 7.2 GHz, and 0\farcs081 $\times$ 0\farcs063 at 32.5 GHz.}
\end{deluxetable*}

In Table \ref{tbl-1}, we report the flux density of the entire system, each nucleus, and the resulting 
spectral indices. We used a curve of growth method to derive the flux densities. For the integrated 
flux density, we progressively $(u, v)$ tapered and re-imaged the data, recording the total flux density 
above a signal-to-noise of 5 at each resolution. These flux densities agree with those measured from the
imaging of individual arrays (see Section \ref{sec:obs}). For the flux densities of the individual nuclei, we 
used CASA's \texttt{imstat} task to place circular apertures around each component, varying their radii. 
We plotted the flux density against aperture radius and looked for convergence in this curve-of-growth to 
identify the true flux density. We also independently measured the integrated flux of the south-west 
component seen at C band (see Figure \ref{fig1}a), using an aperture in the CASA \texttt{viewer}, and 
found that it encloses $\sim$~3\% of the total 5.95 GHz flux density. The integrated flux densities at both 
frequencies agree, within the reported errors, with predicted values based on the modeled integrated 
SED published in the literature \citep[e.g.,][]{Ana00}.

Using these flux densities, we calculated the spectral indices, $\alpha$ for $\mathrm{F_{\nu} \propto \nu^{\alpha}}$, of 
the whole system and each nucleus via

\begin{equation}
\label{eq:index}
\alpha_{1-2} = \frac{\log F_1 - \log F_2 }{\log \nu_1 - \log \nu_2}~,
\end{equation}

\noindent where $F_1$~and $F_2$~are the flux densities at frequencies $\nu_1$ and $\nu_2$. Equation \ref{eq:index} will be
valid for flux densities $\mathrm{F_{1}~and~F_{2}}$ over matched apertures (or for integrated values over whole systems).

\subsection{Comparison to Gas and Recent Star Formation}
\label{sec:calc_snr}

To assess the degree to which the measured sizes are characteristic of the whole system, 
we compared our maps to known distributions of emission from gas and recent radio supernovae
and/or supernova remnants (RSNe/SNRs). 
In Figure \ref{fig1}b, we plot C band contours over the CO (3$\to$2) map of \citet{Sak08} (restoring 
Gaussian beam with FWHM 0\farcs38 $\times$ 0\farcs28 at a p.a $\approx$ 23$^{\circ}$). 

The RSNe/SNRs trace recent star formation, and they can accelerate cosmic ray (CR) electrons 
that emit synchrotron radiation. We built a map of RSN/SNR number density from the locations 
of 49 point sources identified by \citet{Lons06} from 18 cm VLBI observations. On our 33~GHz 
astrometric grid, we convolved delta functions with a fixed, fiducial intensity at the positions of the point 
sources with our 33 GHz beam\footnote{\citet{Lons06} report offsets from the center of Arp 220, 
which we take to be $\alpha_{2000} = \mathrm{15^{h}34^{m}57.259^{s}}$, $\delta_{2000} = \mathrm{23^{\circ}30'11''.409}$.}. 
In Figures \ref{fig1}c and \ref{fig1}d, we compare the 33~GHz map to the distribution of recent RSNe/SNRs.

Note that although we use RSNe/SNRs as signposts of recent star formation, the VLBI sources do not
contribute significantly to the flux that we observe. For a typical synchrotron spectral index $\alpha_{\rm 1.7-33GHz} = -0.7$, the 
\citet{Lons06} RSNe/SNRs would contribute 1.5~mJy at 33~GHz and 4.9~mJy at 6~GHz. This contribution would
only account for $\sim$ 2.5 \% of the total flux density that we observe with the VLA. Even at 18~cm, the \citet{Lons06} RSNe/SNRs 
have integrated flux only $\sim$ 12 mJy at 18 cm, or $\sim$ 4\% of the total flux density of Arp 220 
at that frequency \citep{W&B10}. This contribution is small compared to the 10\% expected fraction 
in normal spiral galaxies like M31 or the Milky Way \citep{Poo69,Ilo72}. This difference is most
likely due to free-free absorption at 18 cm (see Section \ref{sec:od}).

\subsection{The Morphology of Arp 220 nuclei at 33 GHz}
\label{sec:calc_prof}
\begin{deluxetable*}{ccccc}[th]
\tabletypesize{\scriptsize}
\tablecaption{Best-Fit Morphology for the Nuclei of Arp 220 at 33 GHz \label{tbl-2}}
\tablewidth{0pt}
\tablehead{
\colhead{Parameter} & \multicolumn{2}{c}{East nucleus} & \multicolumn{2}{c}{West nucleus}\\
\colhead{}&\colhead{Deconvolved\tablenotemark{a}}&\colhead{Convolved\tablenotemark{b}}&
\colhead{Deconvolved\tablenotemark{a}}&\colhead{Convolved\tablenotemark{b}}
}
\startdata
\cutinhead{\bf{Exponential Disk Model}}
Scale length (pc) & 30.3 $\pm$ 4.6 &  34.0 $\pm$ 5.1 & 21.0 $\pm$ 3.2 & 25.4 $\pm$ 3.8\\ 
Peak intensity (mJy~beam$^{-1}$) &  6.0 $\pm$ 0.7 &  4.8 $\pm$ 0.6  & 13.4 $\pm$ 1.6 &  9.0 $\pm$ 1.1\\ 
Position angle ($\arcdeg$) & 54.7 $\pm$ 0.6 & 55.4 $\pm$ 0.6 & 79.4 $\pm$ 0.8 & 77.3 $\pm$ 0.8\\
Inclination ($\arcdeg$) & 57.9 $\pm$ 0.6 & 55.4 $\pm$ 0.6 & 53.5 $\pm$ 0.5 & 49.1 $\pm$ 0.5\\
$\mathrm{R_{50}}$ (pc)\tablenotemark{*} & 50.8 $\pm$ 7.6 & 57.0 $\pm$ 8.6  & 35.2 $\pm$ 5.3 & 42.7 $\pm$ 6.4\\
\cutinhead{\bf{Two Dimensional Gaussian Fitting}}
FWHM major axis (pc) &85.9 $\pm$ 8.6 & 90.8 $\pm$ 9.1 & 63.7 $\pm$ 6.4 & 70.2 $\pm$ 7.0\\
FWHM minor axis (pc) &46.3 $\pm$ 4.6 & 51.8 $\pm$ 5.2 & 38.0 $\pm$ 3.8 & 44.7 $\pm$ 4.5\\
Position angle ($^{\circ}$) &56.0 $\pm$ 1.1 & 56.5 $\pm$ 1.1 & 78.7 $\pm$ 1.6 & 77.1 $\pm$ 1.5\\
\cutinhead{\bf{Observed Half-Light Radius \tablenotemark{c}}}
$\mathrm{R_{50sky}}$ (pc) &  & 73.3 $\pm$ 7.5 & & 45.5$\pm$ 3.7\\
\enddata

\tablenotetext{a}{Parameters that construct the best image, compared to the observed one, 
after convolving the model with the reported clean beam (see Section \ref{sec:calc_prof} for details).}
\tablenotetext{b}{Best fit parameters that reconstruct the observed image without accounting for the beam.}
\tablenotetext{c}{Taking $\mathrm{R_{50sky} = \sqrt{A_{50sky}/(\pi~\cos i)}}$, where $i$ is the inclination 
obtained from the exponential disk model and $A_{\rm 50sky}$ is the observed area 
enclosing half of the total 33 GHz flux density. The effects of the beam are not accounted for in this size metric.}
\tablenotetext{*}{This is an analytical solution obtained by using the scale length parameter from the model. We refer to the deconvolved column as ${\rm R_{50d}}$.}
\tablecomments{The reported parameters were obtained by fitting a 2-D exponential and Gaussian distribution, respectively. The quoted uncertainties
reflect the systematic uncertainty from varying the goodness of fit statistic or other methodology in the fit. In the case of the peak intensity, the uncertainty is determined 
by the flux density calibrator error, ~12\%. In all 
cases, the errors from the fit are negligible. For Arp 220 ($d_{L}$ = 77 Mpc), 10 pc $\approx$ 0\farcs03 or 0\farcs1 = 36.9 pc }

\end{deluxetable*}

The smooth, ellipsoidal isophotes in Figure \ref{fig1} suggest a disk-like geometry. We modeled 
the 3 $\sigma$ clipped image of Arp 220 at 33 GHz (outermost contour in Figure \ref{fig1}c) using 
a 2-D non-linear least-squares fitting technique. 
After experimenting with Gaussian, exponential, S\'{e}rsic and hybrid profiles, we found that the two 
disks are reasonably described by thin, tilted exponential disks. We fit both nuclei simultaneously 
by varying, without constraints, the amplitude, position angle (p.a.), inclination, center, and scale length of each nucleus. 
Although the parameters did not have constraints, the starting points were educated guesses of the final 
parameters. In each case, we construct the model image, convolve it with the synthesized beam of our 
observations, and compare the model and observed intensities to derive $\chi^2$. Note that the results
for the inclination of the disks represent lower limits because we assume thin disks. 

The best fit parameters from the model fitting, along with associated
uncertainties, are reported in Table \ref{tbl-2}. In addition to
deriving formal uncertainties, we gauge the accuracy of our fit by
varying our approach among several reasonable methods. For example, we
adopt a logarithmic, rather than linear, goodness of fit statistic and
we fit the radial profile rather than the image itself. These imply an
uncertainty of $\approx 15\%$ for the scale length and a few percent
for p.a. and inclination. The error in the normalization is dominated
by our overall uncertainty in the amplitude calibration ($\approx
12\%$). As another point of comparison, we also report the results of
simple Gaussian fitting, although, we emphasize that the residuals are
substantially poorer for this approach at low and high radius. 

From the exponential model, we obtain deconvolved scale lengths of
$30$ and $21$~pc for the east and west nucleus, respectively. Our
results for p.a., inclination, and center did not vary significantly
with the choice of functional form. While the fits appear to be good
descriptions, they are not perfect. From the residual images, we found
that the western nucleus showed higher residuals in the disk than the
eastern nucleus, while the center of the eastern nucleus had higher
residuals than the western nucleus.

\begin{figure*}[tbh]
\includegraphics[width=0.51\textwidth]{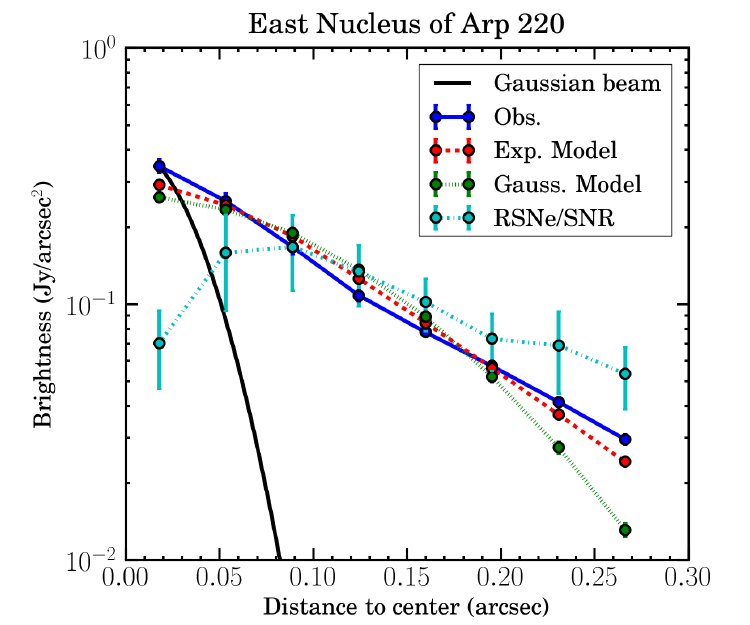}
\includegraphics[width=0.51\textwidth]{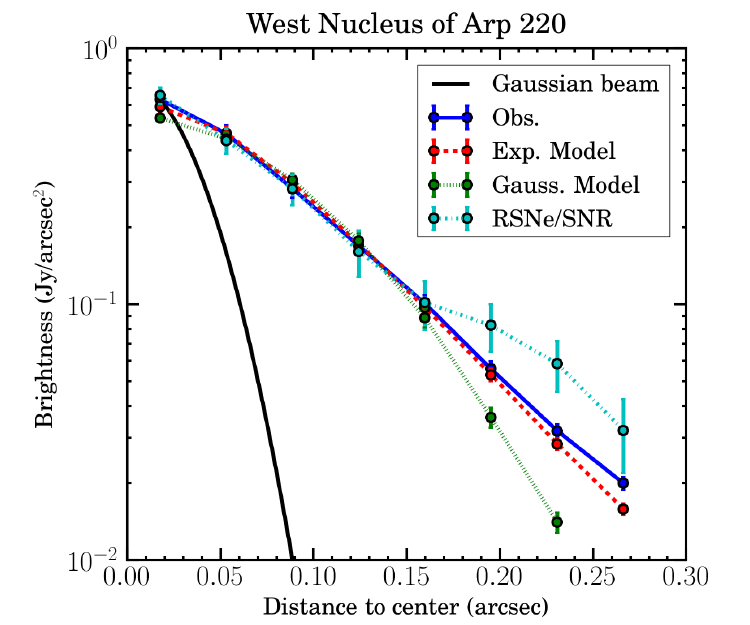}

\figcaption{Azimuthally averaged radial profiles for the deprojected image of the east nucleus 
{\em (left)} and west nucleus {\em (right)} in semilog space. Blue solid curves are 33 GHz 
emission, and cyan dotted dashed line curves are the RSN/SNR number density profiles scaled to match
the 33 GHz radial profile at 0\farcs09. Red dotted lines and green dashed lines are the radial profiles of the exponential 
and Gaussian model images, respectively, convolved with the beam. The black solid lines represent the Gaussian beam radial profile. \label{fig2}}
\end{figure*}

Both nuclei are well resolved, showing significant extent compared with the synthesized beam. The implied 
deconvolved half-light radii, ${\rm R_{50d}}$, are $51$ and $35$~pc, respectively; that is, if viewed face-on, we would 
expect half the emission from Arp 220 to come from nuclear disks $\sim 100$ (east) and $\sim 70$ 
(west) pc across. In Figures \ref{fig1} and \ref{fig2}, we have also shown that the size measurement 
agrees with that implied by the RSN/SNR distribution \citep{Lons06}.

The p.a.'s~of the east and west nuclei agree well with those of the kinematic major axes of the disks 
measured from sub-millimeter CO observations \citep{Sak99,Sak08}, H I absorption observations 
from \citet{Mund01}, and $\mathrm{H53\alpha}$ radio recombination line \citet{RoRi05}. The velocity 
gradients along these position angles on individual nuclei have also been observed in the 2~$\mu$m 
$\mathrm{H_{2}}$ line \citep{Gen01} which traces hot molecular gas and 2.3~$\mu$m CO absorption 
(the latter traces stellar velocities: \citep{En11}).

We performed two checks on the size measurements. First, as a point of
comparison, we report in Table \ref{tbl-2} a 2-D Gaussian fit to each
nucleus at 33~GHz. We obtained deconvolved FWHM sizes of 0\farcs23
$\times$ 0\farcs13 ($\mathrm{86~\times~46~pc^{2}}$) for the eastern
and 0\farcs17 $\times$ 0\farcs10 ($\mathrm{64~\times~38~pc^{2}}$) for
the western nucleus. These sizes agree fairly well with previous,
marginally resolved, estimates at other frequencies. \citet{DE07}
found a deconvolved major axis size of 0\farcs19 = 70~pc for the
western nucleus at 1.3 mm. \citet{Sak08} found major axis sizes of
0\farcs27 = 100~pc (FWHM, east) and 0\farcs16 = 59~pc (FWHM, west) at
860~$\mu$m. However, we show through radial profiles (Figure
\ref{fig2}) that the disks are better described by an exponential
morphology. In fact, the deconvolved Gaussian fit would
underestimate the deconvolved half-light diameter of the disks by
9 \%(west) and 15\% (east), if we account for inclination effect in the 
gaussian fit, i.e, the deconvolved $\mathrm{FWHM_{major}}$ is smaller by 
$\sim$~9\% and 15\% when compared to the deconvolved half-light diameter ($2
\times {\rm R_{50d}}$). 

Second, we calculated the area on the sky containing half of the flux associated with each nucleus. 
This very basic measure still suffers from beam dilution and inclination effects, but provides a 
measure of size that is independent of the functional form. We derived this image-based $A_{\rm 50sky}$ by 
identifying the isointensity contour that encloses 50\% of the total flux density of each nucleus. 
We summed the area of the pixels (pixel size = 0\farcs02) enclosed within that contour and 
estimate the observed radius for $i=55.4\arcdeg$ (east)
and $i=49.1\arcdeg$ (west), and $\mathrm{R_{50sky} = \sqrt{A_{50sky}/(\pi~\cos i)}}$. This is $\sim$~73 
pc for the eastern nucleus and $\sim$~46 pc for the western nucleus. We report the observed ${\rm R_{50sky}}$ 
values in Table \ref{tbl-2}. As with the Gaussian, the measured area shows broad agreement with 
the exponential profile fitting, though differing in detail.

We created deprojected, azimuthally averaged radial profiles of 33~GHz intensity to assess the 
accuracy of our models and compare the structure of the nuclei to that of the RSN/SNR number 
density map. Assuming a thin tilted ring geometry, we calculated deprojected profiles for the observed
emission, the emission in the convolved model, and the RSN/SNR number density map. In each case, the 
center of the profiles correspond to the highest intensity pixel in the observed image for each nucleus.
We plot these profiles in Figure \ref{fig2}. 
We included only emission above a signal-to-noise ratio of 3 and then averaged the intensity in 
a series of inclined, $\sim$0\farcs035-wide (half the clean beam size) rings, adopting the best-fit 
model inclination and p.a. from our modeling (Table \ref{tbl-2}). Note that these rings oversample 
the $\sim$ 0\farcs07 beam, so that adjacent bins in Figure \ref{fig2} are not independent. We 
normalized the RSN/SNR radial profile to match the 33~GHz profile at $r \approx 0\farcs09$. 
The plotted error bars were calculated from the standard deviation of the flux within each annulus 
divided by the square root of the area in that annulus expressed in units of the beam size 
(i.e., the number of independent beams).

In Figure \ref{fig2}, we show that the exponential model matches the data well for both nuclei, 
matching slightly better for the western nucleus. Meanwhile, the Gaussian profile is not as good 
as the exponential profile when compared to the observed data, being particularly poorer in the 
outer parts of the disks. The linearity of the semilog profiles also confirms (and motivates) our adoption 
of an exponential functional form. The RSN/SNR radial profiles mostly follow the integrated radio emission profiles 
(and thus also the model). The agreement is better in the western nucleus. The eastern nucleus 
lacks a bright central peak and shows a somewhat more scattered distribution, which could be caused 
by stochasticity and timescale effects; i.e., there just may not be enough RSNe/SNRs visible to give a 
smooth appearance (compare to Figure \ref{fig1}d to see the clumpy nature of the SNe distribution). The same
idea applies for the outskirts of the western nucleus. In the rest of the paper, we will consider that the RSN/SNR 
number density distribution follow the continuum emission observed at 33 GHz closely enough that we can
take our measured 33~GHz sizes as indicative of the distribution of active star formation in Arp 220.

\subsection {Brightness Temperatures}

\label{sec:Tb}
\begin{deluxetable}{cccccc}
\tabletypesize{\scriptsize}
\tablecaption{Average Brightness Temperatures in Arp 220 \label{tbl-3}}
\tablewidth{0pt}
\tablehead{
\colhead{Frequency} & \colhead{Total system} & \colhead{East nucleus} & \colhead{West nucleus}\\
\colhead{(GHz)}&\colhead{Log[$T_{b}$(K)]}& \colhead{Log[$T_{b}$(K)]}& \colhead{Log[$T_{b}$(K)]}
}
\startdata
5.95  & 4.64 $\pm$ 0.07 & 4.43 $\pm$ 0.10 & 4.81 $\pm$ 0.10\\
32.5  & 2.7 $\pm$ 0.1 & 2.5 $\pm$ 0.1 & 2.9 $\pm$ 0.1
\enddata
\tablecomments{Integrated values are calculated within ${\rm A_{50d}}$ (deconvolved modeled size) 
at 33 GHz (more in Section~\ref{sec:Tb}). The uncertainties follow from propagation of uncertainties
quoted earlier in this paper.}
\end{deluxetable}

The brightness temperature, ${\rm T_{b}}$, can be used to constrain the emission mechanism 
and energy source, and may give clues regarding the optical depth. With well resolved sizes, 
we can circumvent beam dilution that often confuses estimates of ${\rm T_{b}}$. We calculated ${\rm T_{b}}$ 
using the Rayleigh-Jeans approximation via

\begin{equation}
\label{eq:Tb}
T_{b} = \Big(\frac{S_{\nu}}{\Omega_{source}}\Big)\frac{c^{2}}{2 k_{B} \nu^{2}}~,
\end{equation}
\noindent with $S_{\nu}$ the flux density at frequency $\nu$ and $\Omega_{source}$ the area subtended by the source.

We report average Rayleigh-Jeans brightness temperatures, $T_{b}$, in Table \ref{tbl-3}. From our model,
we take the area $\mathrm{A_{50d} \equiv \pi R_{50d}^2}$ that we expect to enclose half the emission if the system 
were viewed face on (see the ``deconvolved'' column in Table \ref{tbl-2}). Assuming this to be the true area of the disk at all frequencies, 
we derive average ${\rm T_{b}}$ over the half-light region. This means we used half of the observed flux density for ${\rm S_{\nu}}$ and 
${\rm A_{50d}}$ for ${\rm \Omega_{source}}$ in Equation \ref{eq:Tb}. Our rationale for this assumption is that the 33~GHz image
appear to be optically thin, high-resolution tracer of the distribution of recent star formation. Assuming
that this structure is common across wavelength regimes allows us to use a ``true'' size in place of a size observed
with a much coarser beam. We also calculated the peak ${\rm T_{b}}$ at each band from the peak flux density and the 
area of the clean beam at each frequency. This peak ${\rm T_{b}}$ is higher at 33 GHz than at 6 GHz;
this simply reflects that the area of Arp 220 at 33 GHz is smaller than the beam size at 6~GHz. For exactly this reason --- the small size 
of Arp 220 and the variable resolution at different frequencies --- the peak measurement has limited 
utility and we only report the ``average'' version.

The average ${\rm T_{b}}$ is from $10^{4.4}$~K at 6 GHz to $10^{2.5}$ ($\approx 300$)~K at 33 GHz, for the east nucleus and 
$10^{4.8}$~K at 6 GHz to $10^{2.9}$ ($\approx 800$)~K at 33 GHz, for the west nucleus.

\section{Discussion}
\label{sec:disc}

Figure \ref{fig1} shows that our observations clearly separated the nuclei at both 6 and 33 GHz and resolve the structure
of both nuclei is resolved at 33 GHz. We find a projected nuclear separation of 0\farcs96 $\pm$ 0\farcs01 
(354 $\pm$ 4 pc), in agreement with previous works \citep[e.g.,][]{Sco98,Soi99,RoRi05,Sak08}. In the following 
subsections, we discuss the radio continuum emission processes, the correspondence with gas and dust emission, 
and consequences of the small sizes of the emission regions. There is a large scale agreement between the 
locations of radio continuum emission, CO, and RSNe/SNRs, suggesting that the sizes of the 
radio continuum sources may be viewed as characteristic of the system.

\subsection{Synchrotron Produces Most of the 33 GHz Emission}
\label{sec:synch}

Synchrotron radiation appears to produce most of the continuum emission at both 6 and 33 GHz. 
The high brightness temperature of a few $\times$ $10^4$~K, inferred at 6 GHz by using the 33 GHz 
nuclear sizes, argues in this direction. This high brightness temperature cannot come from {\sc Hii} regions,
even if they are completely opaque, because in a purely thermal environment, the electron temperature 
of such regions should not exceed $10^4$~K. If we combined the high brightness temperature 
with the observed internal C band spectral index of the total system, $\alpha_{\rm 4.7-7.2GHz} = -0.61 \pm 0.04$, 
we infer that most of the emission at 6 GHz is synchrotron.

The spectral index between 6 and 33 GHz, $\alpha_{\rm 6-33GHz} = -0.69 \pm 0.05$, matches the 
internal C band $\alpha$ of the total system within the errors; the same is true for the two nuclei 
separately (see Table \ref{tbl-1}). The similarity between $\alpha_{\rm 4.7-7.2GHz}$ and $\alpha_{\rm 6-33GHz}$ indicates 
no significant spectral flattening between 6 and 33 GHz, suggesting that synchrotron dominates 
the emission across this range of frequencies. Reinforcing this point, our total flux density and 
spectral index agree with the predictions made by \citet{Ana00} (see their Figure 10b). They found 
synchrotron emission to dominate below $\sim 60$ GHz, and estimated the thermal fraction at 
6 GHz and 33 GHz to be $\sim$~15\% and 35\%, respectively. If we assume a non-thermal spectral
index of $-0.76$ \citep[see Table 9 in][]{Ana00} and a typical thermal spectra index of $-0.1$, we obtain 
$\alpha_{\rm 6-33GHz} \approx -0.62$ and $\alpha_{\rm 4.7-7.2GHz} \approx -0.65$, which is consistent 
with what we measure and with our observation of a constant spectral index from 6 to 33 GHz. 
This result is also consistent with previous results showing lower thermal fractions at 33 GHz for merging 
starbursts compared to normal galaxies \citep{Mur13}.

The overall star formation rate of the system provides an alternate way to estimate the 
expected thermal radio continuum emission. Beginning with the IR (8 to 1000 $\mu$m) luminosity of 
Arp 220, we estimate the expected thermal luminosity of the system if the IR is all 
due to star formation by following star formation rate conversions from Table 8 in 
\citet{Mur12}\footnote{\citet{Mur12} uses a Kroupa IMF to derive the theoretical SFR conversions. The operation
is equivalent to using such an IMF to relate the ionizing photon production (traced by thermal radio emission) 
to bolometric luminosity (traced by IR).} and assuming an electron temperature of 7500~K \citep{Ana00}. This approach 
predicts a thermal fraction of $\sim$ 55\% at 33 GHz and $\sim$ 20\% at 6 GHz. This is in good agreement with the thermal fraction
at 33 GHz obtained in \citet{Con92} for a prototypical starburst, M82. However, if we derive the expected 
spectral index as we did at the end of the previous paragraph (assuming 55\% of thermal fraction at 33 GHz), 
we obtain $\alpha_{\rm 6-33GHz} \approx -0.42$ and $\alpha_{\rm 4.7-7.2GHz} \approx -0.63$, which 
deviates considerably from what we observe between 6 and 33 GHz. Note that $\alpha_{\rm 4.7-7.2GHz}$ does not
vary significantly from what we observe, this is due to the small thermal fraction expected at this frequency range.

The easiest explanation for the lower-than expected thermal flux is that a significant fraction of the 
ionizing photons produced by young stars are absorbed by dust before they produce ionizations. 
This would lower the free-free estimate in the calculation. We estimate that to match 
our observations, we would require that at least 20\% of the ionizing photons be absorbed by dust. 
This number seems plausible for an environment as dust embedded as Arp 220 (see Section \ref{sec:od}) 
and is consistent with some of the arguments made when considering the apparent deficit of IR cooling 
line emission \citep[][and references therein]{DiSa13}. Alternatively, an IMF that produces more 
bolometric light (and thus IR and likely SNe) relative to ionizing photons could resolve the discrepancy. 
That is, we could invoke an ``intermediate-heavy'' IMF compared to that used in \citet{Mur12}. 
We could also reconcile our two estimates if the synchrotron spectral slope decays drastically 
between 6 and 33 GHz, so that the apparent 33-6 GHz index is a combination of very steep, 
curving synchrotron and emerging thermal emission. However, this would require that thermal 
emission make up most of the SED at higher frequencies, which is not observed \citep[see][]{Ana00,Cle10}. 

Our best interpretation of the data is that the 33 GHz emission is mostly synchrotron, in mild contrast 
with a typical starburst galaxy \citep[see Figure 1 in][]{Con92}. We suggest that the most likely cause 
is the suppression of thermal radio emission as dust absorbs ionizing photons.

\subsection{The Radio Emission Coincides with Gas, Hot Dust and RSN/SNR}
\label{sec:SF}

In Figure \ref{fig1}b, we show the 6 GHz emission is largely
co-spatial with CO emission. The 6 GHz emission is our more sensitive
band, with a beam nearly matched to the CO, and --- as just discussed
--- we expect that it traces the same synchrotron emission as the 33
GHz. The CO and 6 GHz emission cover roughly the same area, have
broadly coincident peaks, and both show an extended faint feature to
the southwest \citep[][note a similar coincidence between 18cm
emission and the Arp 220 starburst traced in the
near-IR]{Mazz92}. The distributions of CO and 6 GHz emission do
significantly differ in detail. The ratio of fluxes for the two nuclei
is 1:2 (east:west) for CO and almost 1:1 for continuum. In the west
nucleus, the morphology is more centrally concentrated at 6 GHz
compared to the CO map. Some of these differences may reflect real
differences between the current gas reservoir and recent star
formation, but they may also reflect temperature and optical depth
effects. The CO (3$\to$2) emission in this region shows good evidence
for optical thickness \citep{Sak08}, and the densities are high enough
that the gas temperature will be likely coupled to the dust ($\sim
100$~K).Therefore, making a straightforward interpretation
of the CO in terms of column density is challenging. We draw the broad conclusion from 
Figure \ref{fig1}b that the synchrotron originates from the same region as, and in very rough proportion to,
the molecular gas supply.

A similar situation is also observed on smaller spatial scales by the
comparison of the RSN/SNR number density map to the 33 GHz map (see
Figure \ref{fig1}d). The distributions are co-spatial, but the continuum
map appears smoother than the map made from individual RSN/SNR. This
is particularly evident in the eastern nucleus, where the covering
fraction of VLBA point sources is small --- perhaps a result of
stochasticity in the rate and lifetime of SN visible using VLBI
measurements. The radial profiles in Figure \ref{fig2} highlight the
quantitative agreement between the continuum and the RSNe/SNRs
distribution even more. After azimuthal averaging, the VLBA point
source maps are a fairly close match to the 33 GHz continuum.

Such a close match between the 33 GHz continuum extent and the RSN/SNR number density map 
may not be too surprising: if synchrotron radiation arises from cosmic ray (CR) electrons accelerated by SN shocks, 
then the 33 GHz continuum emission might be expected to resemble a ``puffed up'' version of the RSN/SNR distribution 
due to the diffusion of CR electrons. Instead the distributions match quite well, consistent with most 
of the 33 GHz emission coming from very close to the original RSN/SNR and little diffusion or secondary 
CR electron production. This lack of significant propagation could be explained by the cooling timescales 
being much smaller than the diffusion time. This is expected in compact starbursts with magnetic fields of 
the order of a mG \citep[see measurements from][based on Zeeman splitting of OH megamaser emission]{Rob08,McBri14}, 
like Arp 220 \citep[see Figure 1 in][]{Mur09}, though not in normal galaxies \citep{Mur06}. For 
Arp 220, the cooling time of CR electrons at 33 GHz is $\sim$$\mathrm{10^{3}~yr}$, which is a combination 
of synchrotron, bremsstrahlung, ionization and inverse Compton (IC) losses. We make use of Equation 7 in \citet{Mur09} 
for CR electrons with energies greater than 1 GeV to estimate that the synchrotron emitting electrons at 
33 GHz only have time to propagate about 5 pc, which is about 1/10th of the size that we have measured 
for the nuclei. This short diffusion scale yields a synchrotron image that looks very similar to the sites of 
original CR production (the RSN/SNR) and thus the sites of active star formation. The advantage of the 
VLA continuum in this case is that in exchange for coarser native resolution, we achieve sensitivity to most 
of the flux and spatial scales of interest (and potentially still probe a longer timescale).

The similarity of 6 GHz, 33 GHz, CO surface brightness, and the
RSN/SNR number density distributions lead us to view our 33 GHz
measurement as indicative of the true size of the main disks of star
formation and, presumably, gas and hot dust. These morphologies are
also consistent with the nuclear morphologies measured in mid-IR with
the Keck telescope \citep{Soi99}. Perhaps surprisingly, the two disks
appear fairly similar in terms of profile, scale length, and observed
flux. The western nucleus appears hotter and more compact but the
differences are small factors, not an order of magnitude. The
physical interpretation of such similarities is unclear. Possible
explanations include a similarity in the progenitors, or some ``loss of
memory'' during the process of funneling gas to the center of the
galaxies during the ongoing interaction. 

\subsection{The Nuclear Disks are the Most Extreme Starburst Environments in the Local Universe}

\subsubsection{Gas Surface Densities}
\label{sec:size}

Current best estimates of the dynamical mass per nucleus are $\sim 6
\times 10^{9}$~M$_{\odot}$ within $\sim 100$~pc of each nucleus
\citep{En11}. These values are still uncertain, with $\sim 2 \times
10^{9}$~M$_\odot$ representing a likely lower limit in both nuclei
\citep{En11}. The dynamical mass represents an upper limit on the gas
content. Based on dynamical modeling and CO imaging, \citet{D&S98}
estimated the gas content at $1.1 \times 10^9$ and $0.6 \times
10^9$~M$_\odot$ for the eastern and western nucleus, but embedded in a
larger gas disk with total mass $\sim 3 \times 10^9$~M$_\odot$
\citep[see also][]{Sak99,Sak08,DE07}. These estimates mix dynamical
modeling with observations of low-J CO line (up to J = 3$\to$2)
measurements that are likely very optically thick in Arp
220. \citet{Pap12} provide an alternative, but unresolved, estimate by
focusing on higher J CO transitions and high critical density tracers
(e.g., HCN) to estimate a total molecular gas mass of $\mathrm{\sim(15 - 28) 
\times 10^9~M_\odot}$ for the entire system. The difficulty
with this estimate is apportioning this gas mass to the various
components of the system. We consider a conservative approach to be
the following: we assume that half of the total molecular gas mass is
equally distributed between the two nuclei and the other half in an
outer disk \citep[e.g., see][for evidence of an outer disk]{Sak99}.
This implies $\sim 3.5$--$7 \times 10^9$~M$_\odot$ of gas per nucleus
from \citet{Pap12}. This remains in moderate tension with the
dynamical masses because it would imply very high gas fractions, but
given the mismatch in scales (the dynamical masses are estimated on
$\sim 100$~pc scales) and uncertainties in modeling, a factor of $\sim
3$--$5$ uncertainty seems plausible \citep{Sak08}.

The areas that we measure for the Arp 220 nuclei
are stunningly small, especially when compared to the integrated
properties of the system. We adopt the literature gas mass of the
nuclei as $3.5^{+3.5}_{-1.5} \times 10^9$~M$_\odot$, with the lower
bound set by the \citet{En11} values, the upper bound set by
\citet{Pap12}, and the best estimate consistent (with modest tension)
with the latter. We further expect half of the gas mass of each to be
distributed within the half-light deconvolved, face-on, area (${\rm A_{50d}}$) of our
radio images (these trace star formation, so we implicitly assume that
gas and star formation track one another within the system). We thus
compare $\sim 1.75^{+1.75}_{-0.75} \times 10^9$~M$_\odot$ to our
half-light areas to estimate average, nuclear, gas surface densities. The deprojected,
deconvolved gas surface densities are $2.2^{+2.1}_{-1.0} \times 10^5$~M$_\odot$~pc$^{-2}$
(east) and $4.5^{+4.5}_{-1.9} \times 10^5$
(west)~M$_\odot$~pc$^{-2}$. These translate to an average, nuclear, total hydrogen
column densities of $2.7^{+2.7}_{-1.2} \times 10^{25}$~cm$^{-2}$ (east) and
$5.6^{+5.5}_{-2.4} \times 10^{25}$~cm$^{-2}$ (west) (divide these
numbers by 2 for H$_2$ column densities). These nuclear hydrogen columns
are $\sim 3$--$4$ orders of magnitude higher than those derived from X-ray
observations \citep[e.g.,][]{Clet02,Iwa05}, but they roughly agree with those derived 
from observations at $860~\mu$m \citep{Sak08} and $434~\mu$m \citep{Wil14}.
The gas surface densities that we derive roughly resemble the maximum stellar surface density of $\sim 10^5$~M$_\odot$~pc$^{-2}$
found in a compilation of literature data by \citet{Hop10}. Given the large uncertainty in our mass estimate 
(and the scatter in the \citet{Hop10} compilation) Arp 220 appears consistent with producing such a ``maximal" 
stellar surface density system. This is especially true when one considers that feedback and further evolution 
of the system may reduce the efficiency (final fraction of gas converted to stars) in the nuclei below unity 
(a factor of $\sim 1/3$ would produce excellent agreement).

We do not know the thickness of the disks, but by adopting a
spherical geometry we can calculate a lower limit to the H$_2$
particle densities in the nuclei\footnote{The correction to obtain 
the mass inside a sphere of radius
${\rm R_{50d}}$ is larger than the areal correction. For simplicity, we
adopt the correction appropriate for a Gaussian, so that the mass
within ${\rm R_{50d}}$ is $\approx 1/3.4$ of the total mass.}. This is
$3.8^{+3.8}_{-1.6} \times 10^4$~cm$^{-3}$ (east) and $11^{+12}_{-4.5} \times
10^4$~cm$^{-3}$ (west). For comparison, a typical Milky Way
molecular cloud has a surface density $\sim 100$~M$_\odot$~pc$^{-2}$
(N(H)$\sim 10^{22}$~cm$^{-2}$) and average particle density ${\rm n_{H_2}
\sim 100~cm^{-3}}$. In addition to faster free fall times, correspondingly more
efficient star formation, and phenomenal opacity, potential implications of such high molecular gas
densities would include the secondary production of CR electrons and
confinement of CR electrons.

\subsubsection{Infrared Surface Densities and Star Formation Rates}
\label{sec:ir}

By following the same approach, we assume the infrared emission in Arp 220 is 
coincident within our measured radio distribution and explore the implications.
Conventionally, the infrared luminosity surface density, $\Sigma_{\rm IR}$, is defined as the luminosity
per unit area of the system. We calculate $\Sigma_{\rm IR}$ by assuming that half of the infrared luminosity 
(from 8--1000 $\mu$m) is generated within the deconvolved, face-on, half-light area (${\rm A_{50}}$), 
i.e, we calculate an average infrared luminosity surface density, within the half-light area, via

\begin{equation}
\label{eq:edd}
\Sigma_{\rm IR} = \Big(\frac{0.5 \times L_{IR}[8 - 1000\mu m]}{A_{50}}\Big) = \Big(\frac{L_{50}}{A_{50}}\Big)~.
\end{equation}

\noindent Following our measurements above, we use $\mathrm{A_{\rm 50d} \equiv \pi (R_{50d,east}^2+R_{50d,west}^2)}$
to derive a {\it total (face on)} infrared luminosity surface density of $\mathrm{\Sigma_{IR} \sim 6.0^{+2.3}_{-1.5}
\times 10^{13}~L_{\odot}~kpc^{-2}}$ \footnote{The uncertainties in this value, and in the rest of this section, 
correspond to the errors associated to ${\rm R_{50d}}$.}. If we further assume that the ratio of the fluxes between the 
east and west nuclei at 33 GHz ($\sim$ 1:1) holds at infrared 
wavelengths, then using the derived radio $\mathrm{A_{50d} \equiv \pi R_{50d}^2}$ for the individual disks, we obtain 
$\mathrm{\Sigma_{IR} \sim 4.2^{+1.6}_{-0.7} \times 10^{13}~L_{\odot}~kpc^{-2}}$ and 
$\mathrm{\Sigma_{IR} \sim 9.7^{+3.7}_{-2.4} \times 10^{13}~L_{\odot}~kpc^{-2}}$ for the east and west nucleus, respectively.
These values  are more than an order of magnitude higher than those for the central 0.3 pc of the Orion nebula complex and M 82
($\mathrm{\sim 2 \times 10^{12}~L_{\odot}~kpc^{-2}}$ and $\mathrm{\sim 9 \times 10^{11}~L_{\odot}~kpc^{-2}}$, 
respectively \citep{Soi00}), but are closer to those found in the brightest clusters within starburst galaxies 
($\mathrm{\sim 5 \times 10^{13}~L_{\odot}~kpc^{-2}}$ \citep{Meu97}). 
Our estimated surface densities are consistent with \citet{Soi00}, who estimated
infrared luminosity surface densities of $\mathrm{1-6 \times 10^{13}~L_{\odot}~kpc^{-2}}$ based on
mid-IR Keck observations and radio data from \citet{Con91}.

The deprojected SFR surface density (defined as the SFR per unit area in the disk), $\Sigma_{\rm SFR}$, is a close
corollary of the IR luminosity surface density. We estimate this quantity within the 
deconvolved, face-on, ${\rm A_{50d}}$ for each nucleus at 33 GHz, using a 1:1 ratio between east and west,
the radio luminosity to SFR conversion from Table 8 in \citet{Mur12}, and an electron temperature
and non-thermal spectral index of 7500~K and 0.76 (the spectral index 
in \citet{Mur12} is defined with the opposite sign compared to our definition), respectively, from \citet{Ana00}.
We obtain a $\Sigma_{\rm SFR}$ of $\sim$~$10^{3.7 \pm 0.1}$ and $10^{4.1 \pm 0.1}$ $\mathrm{M_{\odot}~yr^{-1}~kpc^{-2}}$ 
within the half-light of the eastern and western nuclei, respectively (divide these
numbers by two to take into account both sides of the disks).
The total SFR calculated from ${\rm L_{IR}}$, 180 $\mathrm{M_{\odot}~yr^{-1}}$, and from the total radio flux density at 
33 GHz (${\rm L_{Radio}}$), 195 $\mathrm{M_{\odot}~yr^{-1}}$, differ by only $\sim$ 10\%, consistent with Arp 220 
lying on the (33 GHz) radio-to-far infrared correlation (and meaning that we would obtain essentially the same $\Sigma_{\rm SFR}$ for
either luminosity). The radio SFR value differs by 20\% from that derived from \citet{Ana00}, which we consider to be within the uncertainties of such calculations.

$\Sigma_{\rm IR}$ tells us coarsely about the density of IR luminosity per unit area, but not necessarily the flux
at the surface of the source, which may have important implications for feedback and depends on the detailed
geometry of the system. A spherical geometry provides a useful limit on the flux at the surface of the system. In this case,
$\mathrm{F_{sphere} = L_{50} / (4~\pi~R_{50d}^2)}$ will be the flux at the surface of a sphere of radius ${\rm R_{50d}}$ with
the luminosity of Arp 220. Using the measured radio sizes, $\mathrm{F_{sphere} \sim 1.5^{+0.6}_{-0.4} \times 10^{13}~L_{\odot}~kpc^{-2}}$ 
for the entire system, $\mathrm{\sim 1.1^{+0.4}_{-0.2} \times 10^{13}~L_{\odot}~kpc^{-2}}$ 
for the eastern nucleus and $\mathrm{\sim 2.4^{+0.9}_{-0.6} \times 10^{13}~L_{\odot}~kpc^{-2}}$ 
for the western nucleus. A less extreme case, one that may well apply to Arp 220, is a two-sided disk. With one half of the
luminosity emergent from each side, we have $\mathrm{F_{disk} = L_{50} / (2~\pi~R_{50d}^2)}$, twice the spherical case.
As one would expect, these values are lower than the simple $\Sigma_{\rm IR}$, but they also differ from one another, 
reinforcing the importance of geometry to the physics of the source\footnote{Optical depth will provide an additional complication.
Here we consider the IR surface brightness near the ultimate source of the luminosity in the region of active star formation. As the 
radiation scatters out of the system, the geometry may change, so that the geometry of the photosphere could differ from the central
source considered here.}.

Yet another subtlety arises in the specific case where one wishes to calculate the flux 
through an area very close to, but just above one side of a disk. This quantity is relevant to the 
often-discussed case of radiation pressure on dust (see Section \ref{sec:edd}) but 
because of projection effects it is not identical to any of the above quantities. In Appendix \ref{app:flux}, we show that this one-sided
flux perpendicular to the disk, which we call $\mathrm{F_{near}}$ is equal to $\mathrm{L_{IR} /(8 \pi R_{50d}^2)}$ (Equation \ref{eq:ap7}) in 
general. This value is further divided by an extra factor of two for the case of the two nuclei of Arp 220 (because $L_{\rm IR}$ combines
the light from the two nuclei). From this calculation, we obtain $\mathrm{F_{near}\sim 1.5^{+0.6}_{-0.4} \times 10^{13}~L_{\odot}~kpc^{-2}}$ for 
the entire system (the same as for F$_{\rm sphere}$, though not for the same reason), $\mathrm{\sim 1.1^{+0.4}_{-0.2} \times 10^{13}~L_{\odot}~kpc^{-2}}$ 
for the east nucleus and $\mathrm{\sim 2.4^{+1.0}_{-0.6} \times 10^{13}~L_{\odot}~kpc^{-2}}$ for the west nucleus. As discussed in
Appendix \ref{app:flux}, these should be the most appropriate fluxes to consider when assessing the impact of pressure from radiation perpendicular to the disk.

The ratio of the flux densities between the east and west nuclei varies with frequency. Some other ratios
for the east:west relation found in the literature include 1:4 at mid-IR \citep{Soi99}, 1:3 at 18 cm \citep[if we 
consider only the contribution of the point sources from][]{Lons06} and 1:2 at sub-mm wavelengths \citep{Sak08}. 
However, most of these observations do not offer high enough spatial resolution to truly isolate the contribution 
of each nucleus. If we ignore this issue and we assume a ratio of 1:4, and we use Equation \ref{eq:ap7},
we obtain $\mathrm{F_{near, east} \sim 4.5^{+1.7}_{-1.1} \times 10^{12}~L_{\odot}~kpc^{-2}}$ and 
$\mathrm{F_{near, west} \sim 3.7^{+1.4}_{-0.9} \times 10^{13}~L_{\odot}~kpc^{-2}}$. 
In every case, we obtain $\mathrm{F_{near} \gtrsim~10^{13}~L_{\odot}~kpc^{-2}}$
for the west nucleus, which is always the higher intensity nucleus.

Note that the same geometric issues discussed here raise a caveat regarding the luminosity of the
source, which was derived under the assumption of isotropic emission \citep{San03,S&M96}.
An optically thick thin disk is not an isotropic emitter. However, neither do our observations constrain 
the geometry of the infrared photosphere, which is the relevant surface for this calculation. We discuss 
this issue in Appendix \ref{app:flux}. Lacking information, we have assumed the isotropic luminosity throughout this
paper, but note the uncertainty \citep[see also][]{DE07,Wil14}. Note that high angular resolution infrared
(from 8--1000 $\mu$m) observations are needed in order to better constrain the true morphology of the IR photosphere of Arp 220.

\subsubsection{Radiation Pressure and Maximal Starburst Models}
\label{sec:edd}

\begin{figure}[tbh]
\includegraphics[scale=0.27]{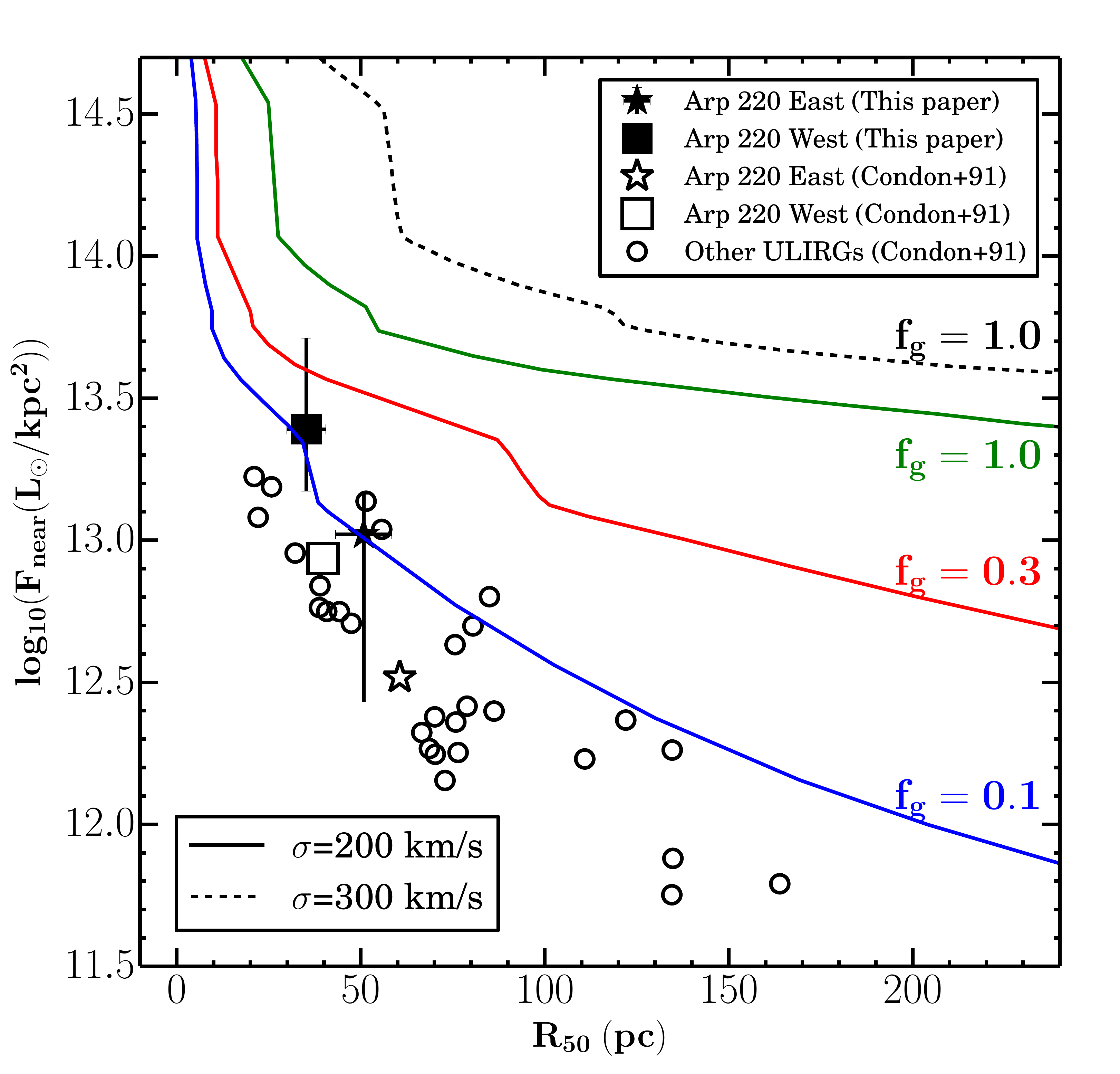}
\centering

\figcaption{Flux versus radius for local ULIRGs from \citet{Con91} ({\it open symbols}), 
and including the values for the eastern ({\it star symbol}) 
and western ({\it square symbol}) nuclei of Arp 220 from this paper ({\it filled symbols}). 
In this figure, we represent a new version of Figure 4 of \citet{Tho05}. The y-axis
corresponds to the flux passing through a surface area near the source, F$_{\rm near}$
(see Appendix \ref{app:flux}), and the x-axis to the deconvolved half-light radius. The solid lines represent the Eddington limits for several
different gas fractions, $\mathrm{f_{g}}$, assuming stellar velocity dispersion of $\sigma$ = 200 km~s$^{-1}$. 
The dashed line shows the model for $\mathrm{f_{g} = 1}$ and $\sigma$ = 300 km~s$^{-1}$. 
From this figure, we observe that Arp 220 appears among the brightest systems but still well
below a conservative dusty Eddington limit described by the green line. \label{fig3}}
\end{figure}

\citet{Sco03} and \citet{Tho05}, argued that for optically thick,
dense starburst galaxies, the critical feedback mechanism acting
against gravitational collapse, and thus star formation, could be radiation
pressure on dust. Although there is ongoing debate about whether or
not radiation pressure on dust represents the dominant feedback
mechanism in compact starbursts \citep[see][for further
discussion]{Kru13,Soc13,Dav14}, the maximal starburst model of
\citet{Sco03} and \citet{Tho05} represents an interesting point of
comparison for our present work\footnote{In addition to radiation
pressure, cosmic ray pressure has been forwarded as a potentially
important feedback mechanism in compact starbursts \citep{Soc08}.}. In Figure
\ref{fig3}, which closely follows Figure 4 of \citet{Tho05}, we
present our new measurements for the flux near the surface of the source
(assuming a thin disk geometry, see Section \ref{sec:ir} and Appendix \ref{app:flux}) in the context of literature
observations and predictions by \citet{Tho05}. The literature
observations show a sample of ULIRGs with sizes based on 8.44
GHz radio maps by \citep{Con91} and luminosities from
IRAS. Following the approach presented in Appendix \ref{app:flux}, we calculate
F$_{\rm near}$ following Equation \ref{eq:ap7} assuming half of the total 
infrared luminosity to be enclosed within $\mathrm{A_{50} = \pi~R_{50}^{2}}$,
where $\mathrm{R_{50}} = b_{\rm maj}/2$. We used the deconvolved FWHM major axis,
$b_{\rm maj}$, from \citet{Con91} in order to account for inclination
effects (we only include resolved sources). Following the discussion in the previous section,
the fluxes for the points in Figure \ref{fig3} differs by a factor of 8 compared to those in
\citet{Tho05}\footnote{This factor of 8 increases for systems having more than one component, in which case
we also divide the total flux among the components. For example, in the case of one individual
region in Arp 299, NGC 3690, the difference is an additional factor of $\sim$ 4 that comes 
from the contribution of that region to the total infrared luminosity of the system \citep{AH00}.
For the other systems having more than one component, we used the relative contribution of each component to
the integrated flux density observed at 8.44 GHz as a template for the relative contribution at infrared wavelengths.}.
for reasons discussed in Appendix \ref{app:flux}.

The error bars of our Arp 220 values (filled points in Figure
\ref{fig3}) correspond to a combination of the uncertainties 
in ${\rm R_{50d}}$, and uncertainty in the contribution of each nucleus to the 
IR luminosity, with the lower/upper limit assuming
a ratio of 1:4 between east and west \citep{Soi99}. To be
conservative, we also assume a 20\% uncertainty in the assumption
that half of the total IR luminosity is coming from ${\rm A_{50}}$.
The plotted values for the flux
correspond to $\mathrm{F_{near, east} \sim 1.1^{+0.4}_{-0.8} \times 10^{13}~L_{\odot}~kpc^{-2}}$
and $\mathrm{F_{near, west} \sim 2.4^{+2.7}_{-0.9} \times 10^{13}~L_{\odot}~kpc^{-2}}$.
In this Figure, we show the ``Eddington'' values for radiation pressure
on dust as solid and dashed lines. These represent an envelope for
which radiation pressure on dust balances self-gravity. As a
result, no equilibrium star-forming system is expected to exist
above this line, hence the ``Eddington limit'' analogy.

The precise value of the limit depends on the size, gas fraction ($\mathrm{f_{g}}$) 
stellar velocity dispersion ($\sigma$), Rosseland mean opacity ($\kappa$), and dust-to-gas ratio of the system, 
leading to the large spread in the model lines seen in the Figure. Our measurements of Arp 220 appear as solid
points in Figure \ref{fig3}. There, the west nucleus of Arp 220 appears among 
the highest brightness systems. However, the west nucleus does not clearly stand out from the other 
ULIRGs with respect to this value, partially because the more compact size means that the 
``maximal'' value is larger for Arp 220 than for larger systems. Overall, all the systems 
plotted in Figure \ref{fig3} lie roughly around the Eddington limit for a f$_{\rm g}$ = 0.1
and $\sigma$ = 200 km s$^{-1}$ disk in \citet{Tho05}, which is indicated by the blue solid line.

If we adopt $\mathrm{f_{g} = 1}$ and $\sigma \approx$ 200 km~s$^{-1}$
\citep[e.g.,][]{Gen01}, indicated by the green line in Figure
\ref{fig3}, then we calculate a conservative Eddington limit of
$\mathrm{\sim 9 \times 10^{13}~L_{\odot} kpc^{-2}}$ for the west
nucleus and $\mathrm{\sim 7 \times 10^{13}~L_{\odot} kpc^{-2}}$ for
the east nucleus. Refined measurements of the geometry, gas fraction, 
opacity, dust-to-gas ratio and kinematics are needed to specify the models 
more precisely. For most plausible assumed disk properties, we can say 
that both Arp 220 nuclei lie well below the \citet{Tho05} Eddington-limited starburst
value, with the western nucleus being the brightest system among the local ULIRGs.
If improved measurements demonstrate one or both nuclei to lie
significantly above this value, one would need to consider luminosity
sources other than star formation \citep[presumably an
AGN][]{Iwa05,DE07,Ran11} or question the basic assumptions about
geometry and equilibrium embedded in the model, but at present little such tension
appears to exist.

Another way to assess the role of radiation pressure in Arp 220 is to assume 
that radiation pressure does represent the dominant force
acting against gravity (see Appendix \ref{app:hydro}) and to calculate the required
gas opacity, $\kappa$, of the system in order to be in hydrostatic equilibrium.
By following Equation \ref{eq:hydro3}, and using the derived values for the gas surface density and
the flux for each nucleus (see Section \ref{sec:size} and \ref{sec:ir}), we find that 
$\mathrm{\kappa_{east} \approx 300~cm^{2}/g}$ and $\mathrm{\kappa_{west} \approx 80~cm^{2}/g}$ would
be required for radiation pressure to balance gravity. Compared to the models from \citet{Sem03}, which were used in the
models from \citet{Tho05}, these appear to be unrealistically high values for the gas opacity. 
We interpret these high values as a reinforcement of our previous findings that the nuclei of
Arp 220 lie below the dusty Eddington limit for $\mathrm{f_{g}}$ $\sim$ 1 and $\sigma$ = 200 km s$^{-1}$ by a factor
of $\sim$ 10, and then are not radiation pressure supported.

\subsubsection{Optical Depth}
\label{sec:od}

The small sizes of the Arp 220 nuclei also imply that optical depth effects will be important across the spectrum, 
even at wavelengths as long as sub-mm. By following the same approach described in Section \ref{sec:Tb}, we 
calculate the average Rayleigh-Jeans brightness temperatures that would be implied by combining the deconvolved 
half-light sizes of the nuclei with half of the 860~$\mu$m continuum flux densities reported by \citet{Sak08} from lower
resolution observations. At this wavelength, we find implied average brightness temperatures of 19~K (east) and 76~K (west), within the deconvolved half-light area
calculated in this paper (${\rm A_{50d}}$), without any accounting for optical depth (which could change the size of the apparent 
emission). In the west nucleus, this approaches the apparent dust temperature of $\sim 90$~K \citep[see][for a more 
thorough idea of the dust temperature in Arp 220]{GoAl12}, so that optical depth must become important by this 
wavelength regime, with $\tau \gtrsim 1$ by 860~$\mu$m in the western nucleus. This neglects any more complicated 
geometric considerations \citep[e.g., see the discussion of an inclined geometry in][]{DE07}, which are likely to make the 
situation even more confused. Optical depth effects appear less severe by $\lambda \sim 1$~mm, combining ${\rm A_{50d}}$ 
with half of the western nucleus dust emission from \citet{DE07}, implies an average $\mathrm{T_{b}^{1mm}} \sim$ 45~K within 
the half-light area.\footnote{This temperature and the 76~K derived from \citet{Sak08} change to 120~K and 215~K, respectively,
if we follow the approach of \citet{DE07}, which uses the full luminosity and defines the size of the source ${\rm \Omega_{source}}$
as $\mathrm{\frac{\pi\theta_{source}}{4~ln(2)}}$ with ${\rm \theta_{source}}$ the geometric mean between the deconvolved FWHM of the 
major and minor axis of the west nucleus. That is, the brightness temperature is higher without accounting for inclination effects.} 
This modest mm optical depth is consistent with measurement of a spectral index steeper than $\alpha=2$ by \citet{Sak08}, 
implying somewhat optically thin emission.

Even this simple calculation demonstrates that by sub-mm wavelengths optical depth effects cannot be neglected, 
especially in the western nucleus. As a result, we would expect high resolution but higher frequency observations, 
e.g., at sub-mm wavelengths with ALMA, to observe a moderately optically thick ``photosphere'' around the galaxy 
and so recover a larger size than we find here \citep[e.g., see the continuum observations
by][]{Wil14}. Similarly, line observations at these wavelengths will need to consider the effects of a moderately 
optically thick sub-mm continuum in their interpretation.

Further, we can estimate the optical depth at 33 GHz. The brightness
temperature calculated for the entire source at 33 GHz is $\sim$ 500~K
(see Table \ref{tbl-3}). The thermal fraction at this frequency is
$\sim 35$\%, so that ${\rm T_{b}}$ of the thermal emission is $\sim$
175~K. The thermal electron temperature (${\rm T_{e}}$) cannot exceed
$\approx 10^{4}$~K because line cooling is high at such high
temperatures. Then taking ${\rm T_{e}\sim10^{4}}$~K and a measured $\sim
175$~K brightness, we estimate the average 33 GHz free-free opacity within
the half-light radius to be $\mathrm{\tau_{Thermal} \sim T_{b}/T_{e} \sim 0.018}$.
Given that $\mathrm{\tau_{Thermal} \propto \nu^{-2.1}}$,
we can calculate that $\mathrm{\tau_{Thermal}\sim1}$ at $\nu$ $\sim$ 5
GHz\footnote{$\tau_{\rm Thermal}$ increases to 0.028 and $\nu$ to 6 GHz if
instead 55\% of the 33 GHz emission is thermal (see discussion in
Section \ref{sec:synch}).}. At 18 cm, $\mathrm{\tau_{Thermal} > 9}$,
which might help explain the low VLBA flux density from
\citet{Lons06}. In addition, as \citet{Sak08} note, the dust opacity
is close to unity at 860 $\mu$m. Thus, the Arp 220 nuclei may be
transparent {\em only} near the middle of the frequency range 5 GHz to
350 GHz. A case can thus be made that the 33 GHz image presented here
is the only existing image that is both optically thin and
resolves the nuclei.

\subsection{Evidence at Radio Wavelengths of a Dominant AGN in the Western Nucleus}
\label{sec:agn}

In our observations, the western nucleus is more compact with a higher ${\rm T_{b}}$ than the eastern nucleus. However, 
consistent with previous VLBI observations, we observe no significant central excess in either Arp 220 images or radial 
profiles (Figures \ref{fig1} and \ref{fig2}). \citet{Parra07} discuss the possibility that one of three VLBI point sources 
showing a flat spectrum ($\alpha > -0.5$), could be an AGN. However, that is one of several possibilities that could explain 
the shape of their spectrum. 

Most of the 33 GHz emission that we observe comes from the compact, but still resolved, disks around the nuclei. 
Specifically, the nuclear beam of the west nucleus contains 20\% of the total flux of the nucleus, while the 
other 80\% arises from the more extended star forming regions (Table \ref{tbl-1}). Our measured ${\rm R_{50d}}$ 
contain similar information (see Table \ref{tbl-2}). We cannot rule out an AGN in the western nucleus, but
if one is present it does not make a dominant, point-like contribution to the overall 33 GHz emission on scales of 
$\approx 30$~pc. Similarly, Arp 220 does not exceed the ``Eddington" value that might eliminate star formation
as a viable power source (see Section \ref{sec:edd}). \citet{Smith98} show that the SN rate and luminosity of Arp 220 are 
broadly consistent with emission only generated by star formation, though uncertainty in the IMF, SN rate, and SFR certainly 
would still allow an AGN contribution.

No high brightness temperature radio core indicative of an AGN is
present. However, given that most AGN are radio-quiet and have weak 
core emission \citep[e.g.,][]{KK89,BB98}, the absence of a radio core does not
rule out the presence of an AGN. Indeed, several studies at other
wavelengths have presented evidence of a possible AGN in Arp 220
\citep[e.g.,][]{Iwa05,DE07,Ran11,I&S14,Wil14}, but to date there is
no clear evidence that the putative AGN makes a significant
contribution to the bolometric luminosity. If our estimate of 
N(H) $\sim$ 10$^{25}$ cm$^{-2}$ in Section \ref{sec:size} is correct,
it would explain why evidence for AGN in Arp 220 has been so elusive. With such high 
column densities any AGN would be Compton thick and undetectable by standard 
AGN diagnostic tools.

\section{Summary}

We present new, high resolution VLA observations of the nearest ULIRG,
Arp 220. Our 33 GHz observations measure the light distribution, which
originates mostly from synchrotron emission, at a wavelength where
optical depth effects are likely negligible. We find exponential
profiles with half-light radii of $51$ and $35$~pc for the eastern and
western nucleus, respectively. The distribution of 33 GHz radio
emission matches the number density distribution of recent RSNe/SNRs
very well. This similarity may result from strong ($\sim$mG)
magnetic fields, which could yield cooling timescales for cosmic ray
electrons that are short compared to the diffusion timescale.
Adopting the measured 33~GHz sizes as characteristic of the
star-forming disks, we derive implied surface densities, H column
densities and volumetric gas densities that strikingly illustrate the
extreme nature of the environment present in Arp 220. Combining our
size measurements with unresolved infrared measurements, we estimate
total fluxes that, although very large, lie well below the conservative predicted values 
for the Eddington-limited ``maximal starburst", though this result is
sensitive to our assumptions. Regardless, the implied
luminosity surface brightness for the west nucleus of Arp 220 is among the most extreme for any
measured system. Given the general uncertain evidence to date of a dominant AGN
in Arp 220, we conclude that the compact size and disk-like morphology
clearly make Arp 220 a prototypical example of the most extreme class 
of star-forming systems in the local universe.

\acknowledgments

{We thank the anonymous referee for helpful comments that made this paper stronger}.
We thank Guillermo Damke for his helpful input in the modeling process, Norman Murray 
for helpful discussions at an early stage of the project, Phil Arras for useful suggestions on Appendix \ref{app:flux}
and Shane Davis for his input, especially in the development of Appendix \ref{app:hydro}, regarding radiation pressure 
and the Eddington limit. We thank the NRAO/UVa star formation 
group (especially Crystal Brogan and Kelsey Johnson) for repeated technical and 
scientific feedback. We also thank Nick Scoville for detailed 
discussions of the astrometry of Arp 220.
A.S.E., G.C.P. and L.B-M. were supported by NSF grant AST 1109475.
L.B-M. was also supported by Fulbright and Becas Chile - CONICYT.
This research made use of the NASA/IPAC Extragalactic 
Database (NED), which is operated by the Jet Propulsion Laboratory, California 
Institute of Technology, under contract with the National Aeronautics and Space
Administration, and NASA's Astrophysics Data System Bibliographic Services. The National 
Radio Astronomy Observatory is a facility of the National Science Foundation operated under 
cooperative agreement by Associated Universities, Inc.

\appendix

\section{Flux Through An Area Just Above an Extended Thin Disk}
\label{app:flux}

Consider a geometrically thin disk with radius $R$ viewed face on. Then consider a small 
area parallel to the disk and at distance $d$ above the disk center. The flux, $F$, passing through the area will be 

\begin{equation}
\label{eq:ap1}
F = \int I cos(\theta) d\Omega~,
\end{equation}

\noindent where $d\Omega = d\phi \sin \theta d\theta$ is the area subtended by an infinitesimal part of the disk. 
The factor $\cos \theta$ accounts for the orientation of the area relative to the patch of emitting disk 
under consideration with $\cos \theta = \frac{d}{\sqrt{R^2+d^2}}$ just as $\sin \theta = \frac{R}{\sqrt{R^2+d^2}}$. 
{\it I} is the specific intensity, which for the optically thick case, is just the source function of the disk and is the same
for all lines of sight (as the disk fills the beam). In the scenario where radiation pressure is important, we consider that
near the disk high optical depth is likely and proceed in the case of Arp 220\footnote{In the optically thin case, {\it I} will depend on the 
path length through the disk, which is larger by a factor of $\cos \theta$ at high viewing angles. 
This factor cancels with the directional $\cos \theta$ in Equation \ref{eq:ap1} so that the optically thin 
case yields a different answer.}.
\noindent The integral in Equation \ref{eq:ap1} goes from 0 to $2 \pi$ in $\phi$ and 0 to $\sin^{-1} (\frac{R}{\sqrt{R^2+d^2}})$ in $\theta$. 
We will immediately change variables so that $x \equiv \sin \theta$ and $dx \equiv \cos \theta~d \theta$. Thus,

\begin{equation}
\label{eq:ap2}
F = I \int_0^{2 \pi} d\phi \int_0^{\frac{R}{\sqrt{R^2+d^2}}} x d x = \pi I~\frac{R^2}{R^2+d^2}~.
\end{equation}

Consider the limit where $d << R$, i.e., where the area the flux is passing through lies just above the disk. Then

\begin{equation}
\label{eq:ap3}
F_{\rm near} =  \pi~I~,
\end{equation}

\noindent similar to the well known relation that the flux at the surface of a blackbody is $\pi B_{\nu}$\footnote {F$_{\rm near}$ will be different by
a factor of two in the optically thin case.}. Similarly, at large $d >> R$, as for an astronomical observation:

\begin{equation}
\label{eq:ap4}
F_{\rm far} =  \pi~I~\frac{R^2}{d^2}~,
\end{equation}

\noindent which is nothing more than the integral of the intensity over the solid angle subtended by the disk. The utility in this calculation
is to relate $I$ back to the luminosity, which for an isotropic emitter is just $\mathrm{L = 4 \pi d^2 F_{\rm far}}$ (see the last paragraph of this appendix for
some caveats regarding this assumption). Then

\begin{equation}
\label{eq:ap5}
F_{\rm far} =  \pi~I~\frac{R^2}{d^2} = \frac{L}{4 \pi d^2}
\end{equation}

\noindent so that,

\begin{equation}
\label{eq:ap6}
I = \frac{L}{4~\pi^2~R^2}~{\rm~and~}~F_{\rm near} = \frac{L}{4~\pi~R^2}~.
\end{equation}

This $F_{\rm near}$ is the flux through a surface near the disk and the departure from the perhaps expected $\mathrm{L/(2~\pi~R^2)}$ is that 
we have included the $\cos \theta$ term in the original setup to account for the projection of the incident intensity onto the unit area.

When considering real observations cast in terms of the half-light area, $A_{50}$ (and recall that the commonly used size at FWHM for a two dimensional Gaussian
is $A_{50}$) an additional factor enters from the fact that $L_{50}$ = $0.5~L$ for $R_{50}$. Then in general:

\begin{equation}
\label{eq:ap7}
F_{\rm near} = \frac{L}{8~\pi~R_{50}^2}~,
\end{equation}

\noindent is the flux that should be used for considering radiation pressure near a large disk 
(where large is defined so that $d << R$ can hold). In Arp 220, an additional factor of two 
comes into play if we assume the luminosity is split between the two nuclei with a ratio 
of 1:1. This decreases the relation to $\mathrm{F_{near} = L/(16~\pi~R_{50}^2)}$ for this specific case. Note
the stark difference, even for the general case, from the commonly adopted $\mathrm{\Sigma_{IR} = L / \pi R_{50}^2}$.

In the case of an {\em optically thick} disk, the assumption of isotropy is not valid and $\mathrm{L \neq 4 \pi d^2 F_{\rm far}}$,
instead $\mathrm{L = 2 \pi d^2 F_{\rm far}}$. Further, the emission is not isotropically distributed, so that if the disk is inclined by an angle, $i$, 
with respect to the line of sight, where $i=0$ is a face-on disk, then $\mathrm{L = 2 \pi d^2 F_{far}/\cos}$ $i$. In other words, the emission comes 
only from the two sides of the disks and an observer finds more flux when the disk is viewed face on (because the constant intensity surface subtends
more solid angle). This is a substantial uncertainty for Arp 220 \citep[e.g., see][]{DE07}: it appears to host two inclined disks and shows good evidence for optical depth
at IR wavelengths. However, we are hesitant to impose any correction to the luminosity in the main analysis because we
we do not know the true geometry of the infrared photosphere, which might very plausibly be more spherical and emit more isotropically than the
nuclear disks picked out by our 33~GHz observations. Therefore throughout the main text we have used the conventional $L = 4 \pi d^2 F_{\rm far}$ but we note this substantial 
uncertainty.

\section{Vertical Hydrostatic Equilibrium for a Simple Radiation Pressure Dominated Disk}
\label{app:hydro}

As a simple check on the plausibility of radiation pressure representing the main means of support, we consider 
vertical hydrostatic equilibrium in a simple gas disk. We consider an infinite slab of surface density $\Sigma$, so that
the integrated weight of the column of gas at the midplane is $\pi G \Sigma^2$ (here $G$ is the gravitational constant). To a 
coarse approximation, radiation pressure can counteract this weight with a pressure set by the momentum flux of photons, $F / 2c$ (where $c$ is the speed 
of light), multiplied by the total opacity of the gas column, $\kappa \Sigma$, where $\kappa$ is the cross section per unit gas mass. Then:

\begin{equation}
\label{eq:hydro1}
%\frac{\pi G \Sigma^2}{2} \sim \frac{\kappa \Sigma F}{c}~.
\pi G \Sigma^2 \sim \frac{\kappa \Sigma F}{2c}~.
\end{equation}

\noindent We can then solve for $\kappa$ in terms of the other properties of the disk and the resulting $\kappa$ expresses the required effective opacity
in order for radiation pressure to balance the weight of the disk:

\begin{equation}
\label{eq:hydro3}
\kappa \sim \frac{2 \pi c G\Sigma}{F}.
\end{equation}

If $F$ and $\Sigma$ are known, comparison of the require $\kappa$ to realistic values represents a zeroth order check of whether radiation pressure represents
a viable support mechanism for a system. As a close corollary, if $\kappa$ is known or can be estimated, assessing the degree to which Equation \ref{eq:hydro1} represents
an inequality offers diagnostic of the importance of radiation pressure to the system.
Note that $\kappa$, as we have written it, will depend on the dust-to-gas ratio, grain properties, overall opacity, and temperature distribution. 
In our simplified thin disk geometry, it does not depend directly on the size of the system but all of these properties may vary substantially as 
a function of disk structure. As a first order approximation, we assume the \citet{Sem03} model values serve as a template of typical values 
that would describe $\kappa$ for the nuclei in Arp 220, but to our knowledge a thorough exploration of $\kappa$ appropriate for the nuclear 
disks of merging galaxies remains lacking in the literature.

\end{document}